\documentclass[twocolumn,preprintnumbers,amsmath,amssymb]{revtex4}
\usepackage{mathrsfs}
\usepackage{amssymb}
\usepackage{graphicx}
\usepackage{dcolumn}
\usepackage{bm}
\usepackage{textcomp}
\usepackage{amsmath}
\usepackage[titletoc]{appendix}

\newcommand\ket[1]{\ensuremath{|#1\rangle}}

\newcommand\oprod[2]{\ensuremath{|#1\rangle\langle#2|}}
\newcommand\mean[1]{\ensuremath{\langle #1\rangle}}

\begin{document}
\title{Measurement-device-independent quantum key distribution with source
state errors and statistical fluctuation}
\author{ Cong Jiang$ ^{1,2}$, Zong-Wen Yu$ ^{1,3}$,
and Xiang-Bin Wang$ ^{1,2,4\footnote{Email
Address: xbwang@mail.tsinghua.edu.cn}\footnote{Also at Center for Atomic and Molecular Nanosciences, Tsinghua University, Beijing 100084, China}}$}

\affiliation{ \centerline{$^{1}$State Key Laboratory of Low
Dimensional Quantum Physics, Department of Physics,} \centerline{Tsinghua University, Beijing 100084,
Peoples Republic of China}
\centerline{$^{2}$ Synergetic Innovation Center of Quantum Information and Quantum Physics, University of Science and Technology of China}\centerline{  Hefei, Anhui 230026, China
 }
\centerline{$^{3}$Data Communication Science and Technology Research Institute, Beijing 100191, China}\centerline{$^{4}$ Jinan Institute of Quantum technology, SAICT, Jinan 250101,
Peoples Republic of China}}
\begin{abstract}
We show how to calculate the secure final key rate in the four-intensity decoy-state MDI-QKD protocol with both source errors and statistical fluctuations with a certain failure probability. Our results rely only on the range of only a few parameters in the source state. All imperfections in this protocol have been taken into consideration without any unverifiable error patterns.

\end{abstract}


\maketitle

\section{Introduction}
Quantum key distribution (QKD) can provide unconditional security based on the laws of quantum physics~\cite{BB84,GRTZ02}. But most existing experiment setups such as imperfect single-photon sources and finite-efficiency detectors are not up to the standard of the theoretical models assumed in the security proof and cause security problems~\cite{PNS,PNS1,lyderson}. Fortunately, the decoy-state method~\cite{ILM,H03,wang05,LMC05,wang06,AYKI,peng,wangyang,rep,njp} can help us assured the security with imperfect single-photon sources~\cite{PNS,PNS1}. Theories including device-independent QKD (DI-QKD)~\cite{ind1} and measurement-device-independent QKD (MDI-QKD)~\cite{curty1,ind3} have been proposed to overcome the finite-efficiency detector problem.

MDI-QKD is based on the idea of entanglement swapping~\cite{curty1,ind3}. There, both Alice and Bob send out quantum signals to the untrusted third party (UTP), but neither of them performs any measurement. The UTP would perform a Bell state measurement for each received pulse pair and announce whether it was a successful event as well as his measurement outcome in the public channel. Those bits corresponding to successful events will be post selected and further processed for the final key. By using the decoy-state method, Alice and Bob can use imperfect single-photon sources~\cite{curty1,wang10} securely in the MDI-QKD. The decoy-state MDI-QKD has been studied both experimentally~\cite{tittel,chan1,liu1,ferreira,tang1,tang2,404km} and theoretically~\cite{wang10,ind2,a1,a2,a3,a4,a5,a6,a7,a8,a9,a10,a11,jiang}.

The existing decoy-state MDI-QKD theory assumes perfect control of the source states. This is an impossible task for any real setup in practice. Refs.~\cite{wangyang,njp} have show how to calculate the secure final key rate with intensity fluctuation, or more generally, the source errors in the Bennett-Brassard 1984 (BB84) decoy-state QKD protocol. In our last work~\cite{jiang}, we get the formulas of secure final key rate with source errors in asymptotic case in decoy-state MDI-QKD protocol without any presumed conditions. (In most cases, we have no idea about the details of source errors, so we cannot make any assumptions to the source error model.) But in real experiments, the total pulse pairs sent out by Alice and Bob are finite and we have to take the statistical fluctuation into consideration. The ideas proposed in Refs.~\cite{a9,a10,a11} such as global optimization and combined constraints greatly enhance the secure final key rate in nonasymptotic case in decoy-state MDI-QKD protocol. In this paper, we will show how to extend the formulas in Ref.~\cite{jiang} to nonasymptotic case with those brilliant ideas. In our method, we have assumed the worst case that Eve knows exactly the error of each pulse. Our result immediately applies to all existing experimental results. Before going further, we emphasize that our results here are unconditionally correct because we have not assumed any unproven conditions. Although there is another approach reported for the issue of intensity error by using the model of attenuation to pulses from an untrusted source, however, there exists counter examples to the elementary equation in that approach, as was shown in the appendix of Ref.~\cite{njp}.

This paper is arranged as follows. In Sec.~\ref{Sec:protocal}, we present the protocol of our method. In Sec.~\ref{secasy}, we show the details of how we extend the formulas in our last work and get the secure final key rate in asymptotic case. In Sec.~\ref{fluctuation}, we show the relationship between the asymptotic values and nonasymptotic values with a fixed failure probability $\xi$. And in Sec.~\ref{simulation}, we present the numerical simulations. The article ends with some concluding remarks.

\section{Protocol}\label{Sec:protocal}
In this paper, we consider the four-intensity decoy-state MDI-QKD protocol. In this protocol, Alice (Bob) has three sources $v_A$, $x_A$, and $y_A$ ($v_B$, $x_B$, and $y_B$) in $X$ basis and one source $z_A$ ($z_B$) in $Z$ basis, and they send pulse pairs (one pulse from Alice, one pulse from Bob) to the UTP one by one. And if the UTP announces that it's a successful event, then we say that the $i$th pulse pair has caused a count. Each pulse sent out by Alice (Bob) is randomly chosen from one of the four sources $l_A$ ($r_B$) with constant probability $p_{l_A}$ ($p_{r_B}$) for $l,r=v,x,y,z$. Sources $v_A$ and $v_B$ are the unstable vacuum sources in $X$ basis. Generally, these pulses from the unstable vacuum sources are not the exact zero photon-number state. $x_A, y_A$ ($x_B, y_B$) are the decoy sources which are used to estimate the lower bound of yield and the upper bound of phase-flip error rate of single-photon pulse pairs. $z_A$ ($z_B$) is the signal source which is used to extract the final key.

We shall use notation $l_Ar_B$ to indicate the two-pulse source when Alice use source $l_A$ and Bob use source $r_B$ to generate a pulse pair. For simplicity, we omit the subscripts of any $l$ and $r$ for a two-pulse source, e.g., source $xy$ is the source in which Alice uses source $x_A$ and Bob uses source $y_B$. We also denote the number of counts and error counts caused by the two-pulse source $lr$ as $N_{lr}$ and $M_{lr}$ respectively. And we use $\mean{N_{lr}}$ and $\mean{M_{lr}}$ as the asymptotic values of corresponding observed values $N_{lr}$ and $M_{lr}$. $N_{lr}$ and $M_{lr}$ are observables and will be regarded as known values.

Suppose Alice and Bob send $N_t$ pulse pairs to the UTP in the whole protocol. In photon-number space, the state of the $i$th pulse pair from source $l_A$ ($r_B$) is
\begin{equation}\label{states}
  \rho_{l_A}^{i}=\sum_{k}a_k^{l,i}\oprod{k}{k} \quad(\rho_{r_B}^{i}=\sum_{k}b_k^{r,i}\oprod{k}{k}),
\end{equation}
for $l,r=v,x,y,z$.

We use superscripts $U,L$ for the upper bound and lower bound of a certain parameter. In particular, given any $k\ge 0$ in Eq.~\eqref{states}, we denote $a_k^{l,U},a_k^{l,L}$ ($b_k^{r,U},b_k^{r,L}$) for the maximum value and minimum value of $a_k^{l,i}$ ($b_k^{r,i}$) for any $i\in N_t$, and $l,r=v,x,y,z$. We assume these bound values are known in the protocol.

\section{Analysis for final key rate in asymptotic case}\label{secasy}
In this section, we discuss how to get the secure final key rate in asymptotic case. We denote the pulse pair in which Alice send out a $\ket{j}$-photon pulse and Bob send out a $\ket{k}$-photon pulse as $\ket{jk}$-photon pulse pair and we denote the number of counts caused by $\ket{jk}$-pulse pairs of source $lr$ as $n_{jk}^{lr}$. Therefore we can formulate the total pulse pair counts caused by source $lr$ by
\begin{equation}\label{Nlr}
\mean{N_{lr}}=\sum_{j,k\ge 0}n_{jk}^{lr}.
\end{equation}
We also need to introduce the following notation
\begin{align}\label{nlr}
\mean{\widetilde{N}_{lr}}=\sum_{j,k\ge 1}n_{jk}^{lr},\quad \mean{n_{lr}^o}=\sum_{j\ge 1}n_{j0}^{lr}+\sum_{k\ge 1}n_{0k}^{lr}+n_{00}^{lr}.
\end{align}
It is easy to know $\mean{N_{lr}}=\mean{\widetilde{N}_{lr}}+\mean{n_{lr}^o}$.

As presented in Ref.~\cite{jiang}, if
\begin{equation}\label{decoycondition1}
 \frac{a_k^{y,L}}{a_k^{x,U}}\ge\frac{a_2^{y,L}}{a_2^{x,U}}\ge\frac{a_1^{y,L}}{a_1^{x,U}},\quad
  \frac{b_k^{y,L}}{b_k^{x,U}}\ge\frac{b_2^{y,L}}{b_2^{x,U}}\ge\frac{b_1^{y,L}}{b_1^{x,U}},
\end{equation}
and
\begin{equation}\label{decoycondition2}
 \frac{a_k^{l,i}}{a_k^{v,i}}\ge\frac{a_1^{l,i}}{a_1^{v,i}}, \quad \frac{b_k^{r,i}}{b_k^{v,i}}\ge\frac{b_1^{r,i}}{b_1^{v,i}},\quad (l,r=x,y),
\end{equation}
hold for all $k\ge 2$, we could formulate the lower bound counts of single-photon pulse pairs as follow:
\begin{equation}
n_{11}^{lr}\ge p_lp_ra_1^{l,L}b_1^{r,L}D_{11},
\end{equation}
where
\begin{equation}\label{D11n}
D_{11}\ge\frac{\frac{a_1^{y,L}b_2^{y,L}}{p_x^2}\mean{\widetilde{N}_{xx}}-\frac{a_1^{x,U}b_2^{x,U}}{p_y^2}\mean{\widetilde{N}_{yy}}}{a_1^{x,U}a_1^{y,L}(b_1^{x,U}b_2^{y,L}-b_2^{x,U}b_1^{y,L})}.
\end{equation}

Our next job is to formulate the upper bound of phase flip error rate. If we denote the error counts caused by $\ket{jk}$-photon pulse pair of source $xx$ as $m_{jk}^{xx}$, we have
\begin{align}
\mean{M_{xx}}=&\sum_{j,k\ge 0}m_{jk}^{xx}=\mean{\widetilde{M}_{xx}}+ \mean{m_{xx}^o}\nonumber \\
\label{Mxx}\ge& \mean{m_{xx}^o}+m_{11}^{xx},
\end{align}
where
\begin{align}
\mean{\widetilde{M}_{xx}}&=\sum_{j,k\ge 1}m_{jk}^{xx},\\
\label{mxx0}\mean{m_{xx}^o}&=\sum_{j\ge 1}m_{j0}^{xx}+\sum_{k\ge 1}m_{0k}^{xx}+m_{00}^{xx}.
\end{align}
Thus we have
\begin{equation}\label{phasefliperrorrate}
e_{11}^{ph}=\frac{m_{11}^{xx}}{n_{11}^{xx}}\le\frac{\mean{M_{xx}}-\mean{m_{xx}^o}}{N_tp_x^2a_1^{x,L}b_1^{x,L}s_{11}^L},
\end{equation}
where $e_{11}^{ph}$ is the phase flip error rate of signal source $zz$ and $s_{11}^L=D_{11}^L/N_t$ is the lower bound of the yield of single-photon pulse pairs.

The most important conclusion we get in Ref.~\cite{jiang} is that we could formulate the lower and upper bounds of $\mean{\widetilde{N}_{xx}}$, $\mean{\widetilde{M}_{xx}}$, and $\mean{\widetilde{N}_{yy}}$ without perfect vacuum source. Actually, we only need the upper bounds of $\mean{\widetilde{M}_{xx}}$ and $\mean{\widetilde{N}_{yy}}$ in this paper. Explicitly, we have
\begin{align}
\label{wideNxx}\mean{\widetilde{M}_{xx}^U}=&\frac{1}{1-\sigma_A^x-\sigma_B^x}\left(
\mean{M_{xx}}-\frac{p_xa_0^{x,L}}{p_va_0^{v,U}}\mean{M_{vx}}\right.\nonumber \\
&\left. -\frac{p_xb_0^{x,L}}{p_vb_0^{v,U}}\mean{M_{xv}}-\frac{p_x^2a_0^{x,U}b_0^{x,U}}{p_v^2a_0^{v,L}b_0^{v,L}}\mean{M_{vv}}\right),\\
\label{wideNyy}\mean{\widetilde{N}_{yy}^U}=&\frac{1}{1-\sigma_A^y-\sigma_B^y}\left( \mean{N_{yy}}-\frac{p_ya_0^{y,L}}{p_va_0^{v,U}}\mean{N_{vy}}\right.\nonumber \\
&\left. -\frac{p_yb_0^{y,L}}{p_vb_0^{v,U}}\mean{N_{yv}}-\frac{p_y^2a_0^{y,U}b_0^{y,U}}{p_v^2a_0^{v,L}b_0^{v,L}}\mean{N_{vv}} \right),
\end{align}
where
\begin{align}\label{sigmaAB}
\begin{split}
\sigma_A^x=\frac{a_0^{x,U}a_1^{v,U}}{a_0^{v,L}a_1^{x,L}},\quad \sigma_B^x=\frac{b_0^{x,U}b_1^{v,U}}{b_0^{v,L}b_1^{x,L}}, \\
\sigma_A^y=\frac{a_0^{y,U}a_1^{v,U}}{a_0^{v,L}a_1^{y,L}},\quad \sigma_B^y=\frac{b_0^{y,U}b_1^{v,U}}{b_0^{v,L}b_1^{y,L}}.
\end{split}
\end{align}

Given the obvious fact that the bit flip error rate must be $50\%$ if the bit is caused by a $\ket{j0}$-photon or $\ket{0k}$-photon pulse pair, we have
\begin{equation}
\mean{n_{xx}^o}=2\mean{m_{xx}^o},
\end{equation}
and we define $\mathcal{H}=\frac{2\mean{m_{xx}^o}}{p_x^2N_t}$ to simplify the following calculation. With Eqs.~\eqref{Mxx} and~\eqref{wideNxx}, it is a pretty easy work to get the lower and upper bounds of $\mathcal{H}$, which is
\begin{equation}\label{mathH}
\frac{2(\mean{M_{xx}}-\mean{\widetilde{M}_{xx}^U})}{p_x^2N_t}\le \mathcal{H}\le \frac{2\mean{M_{xx}}}{p_x^2N_t}.
\end{equation}

To formulate the secure final key rate with the observed counting rates, we need to introduce the following notation
\begin{align}
\begin{split}
\mean{S_{lr}}&=\frac{\mean{N_{lr}}}{p_lp_rN_t}, \quad S_{lr}=\frac{N_{lr}}{p_lp_rN_t}, \\
\mean{T_{lr}}&=\frac{\mean{M_{lr}}}{p_lp_rN_t}, \quad T_{lr}=\frac{M_{lr}}{p_lp_rN_t},
\end{split}
\end{align}
where $S_{lr}$ and $T_{lr}$ are the observed values of the counting rate and error counting rate of source $lr$, and $\mean{S_{lr}}$ and $\mean{T_{lr}}$ are their corresponding asymptotic values. Then we could formulate the final key rate with those asymptotic counting rate values.
With Eqs.~\eqref{wideNxx} and~\eqref{mathH}, we could get the lower and upper bounds values of $\mathcal{H}$
\begin{align}
\label{hu}\mathcal{H}^U=&2\mean{T_{xx}},\\
\label{hl}\mathcal{H}^L=&\frac{2}{1-\sigma_A^x-\sigma_B^x}\left[\frac{a_0^{x,L}}{a_0^{v,U}}\mean{T_{vx}}+\frac{b_0^{x,L}}{b_0^{v,U}}\mean{T_{xv}}\right.\nonumber \\
&\left. -\frac{a_0^{x,U}b_0^{x,U}}{a_{0}^{v,L}b_0^{v,L}}\mean{T_{vv}}-(\sigma_A^x+\sigma_B^x)\mean{T_{xx}}\right],
\end{align}
Given a certain $\mathcal{H}$ $(\mathcal{H}\in[\mathcal{H}^L,\mathcal{H}^U])$, the lower bound of the yield of single-photon pulse pairs is
\begin{equation}
\label{s11}s_{11}^L(\mathcal{H})=\frac{S^+-S^--a_1^{y,L}b_2^{y,L}\mathcal{H}}{a_1^{x,U}a_1^{y,L}(b_1^{x,U}b_2^{y,L}-b_2^{x,U}b_1^{y,L})},
\end{equation}
where
\begin{align}
\label{splus}S^+=&a_1^{y,L}b_2^{y,L}\mean{S_{xx}}+\frac{a_1^{x,U}b_2^{x,U}(\frac{a_0^{y,L}}{a_0^{v,U}}\mean{S_{vy}}+\frac{b_0^{y,L}}{b_0^{v,U}}\mean{S_{yv}})}{1-\sigma_A^y-\sigma_B^y},\\
\label{sminus}S^-=&\frac{a_1^{x,U}b_2^{x,U}(\mean{S_{yy}}+\frac{a_0^{y,U}b_0^{y,U}}{a_0^{v,L}b_0^{v,L}}\mean{S_{vv}})}{1-\sigma_A^y-\sigma_B^y},
\end{align}
and the upper bound of phase flip error rate is
\begin{equation}\label{e11}
e_{11}^{ph,U}(\mathcal{H})=\frac{\mean{T_{xx}}-\frac{1}{2}\mathcal{H}}{a_1^{x,L}b_1^{x,L}s_{11}^L(\mathcal{H})},
\end{equation}
and thus the final key rate is
\begin{align}
\label{RH}\mathcal{R}(\mathcal{H})=&p_z^2\{a_1^{z,L}b_1^{z,L}s_{11}^L(\mathcal{H})[1-H(e_{11}^{ph,U}(\mathcal{H}))]\nonumber \\
&-fS_{zz}H(E_{zz})\},
\end{align}
where $f$ is the error correction inefficiency and $H(x)=-xlog_2x-(1-x)log_2(1-x)$ is the binary Shannon entropy function. And $\mathcal{R}(\mathcal{H})$ is the final key rate of one emissive pulse pair with a certain $\mathcal{H}$. Here we have used the fact that the lower bound of the yield of single-photon pulse pairs in the $Z$ basis can be estimated by its lower bound in the $X$ basis. The secure final key rate is
\begin{equation}
R=min[\mathcal{R}(\mathcal{H})],\quad  \mathcal{H}\in[\mathcal{H}^L,\mathcal{H}^U].
\end{equation}

\section{Analysis for statistical fluctuation}\label{fluctuation}
The formulas of secure final key rate in Sec.~\ref{secasy} are denoted by expected values, but the direct values we get in experiments are observed values. The relationship between expected values and observed values is discussed as follows. Let $X_1,X_2,\dots,X_n$ be $n$ random samples, detected with the value 1 or 0, and let $X$ denote their sum satisfying $X=\sum_{i=1}^nX_i$. $\mu$ is the expected value of $X$. From the Chernoff bound~\cite{tang2,a4,chernoff} with a fixed failure probability $\xi$ which is required in experiments, we have
\begin{align}
\label{mul}\mu^L(X)=&\frac{X}{1+\delta_1(X)},\\
\label{muu}\mu^U(X)=&\frac{X}{1-\delta_2(X)},
\end{align}
where we can obtain the value of $\delta_1(X)$ and $\delta_2(X)$ by solving the following equations
\begin{align}
\label{delta1}\left(\frac{e^{\delta_1}}{(1+\delta_1)^{(1+\delta_1)}}\right)^{\frac{X}{1+\delta_1}}&=\frac{\xi}{2},\\
\label{delta2}\left(\frac{e^{-\delta_2}}{(1-\delta_2)^{(1-\delta_2)}}\right)^{\frac{X}{1-\delta_2}}&=\frac{\xi}{2}.
\end{align}
With Eqs.~\eqref{mul} and~\eqref{muu}, we could get the following useful constraints
\begin{align}
\label{slr}&\mu^L(\mathcal{L}_{lr}S_{lr})\le\mathcal{L}_{lr}\mean{S_{lr}}\le\mu^U(\mathcal{L}_{lr}S_{lr}),\\
\label{txx}&\mathcal{L}_{xx}\mean{T_{xx}}\le \mu^U(\mathcal{L}_{xx}T_{xx}),\\
\label{res1}\begin{split}\begin{cases}
&\mathcal{L}_{xx}\mean{S_{xx}}+\mathcal{L}_{vy}\mean{S_{vy}}\ge \mu^L(\mathcal{L}_{xx}S_{xx}+\mathcal{L}_{vy}S_{vy}),\\
&\mathcal{L}_{xx}\mean{S_{xx}}+\mathcal{L}_{yv}\mean{S_{yv}}\ge \mu^L(\mathcal{L}_{xx}S_{xx}+\mathcal{L}_{yv}S_{yv}),\\
&\mathcal{L}_{vy}\mean{S_{vy}}+\mathcal{L}_{yv}\mean{S_{yv}}\ge \mu^L(\mathcal{L}_{vy}S_{vy}+\mathcal{L}_{yv}S_{yv}),\\
&\mathcal{L}_{xx}\mean{S_{xx}}+\mathcal{L}_{vy}\mean{S_{vy}}+\mathcal{L}_{yv}\mean{S_{yv}}\ge \\
&\quad\quad\quad\quad\quad\quad\mu^L(\mathcal{L}_{xx}S_{xx}+\mathcal{L}_{vy}S_{vy}+\mathcal{L}_{yv}S_{yv}),
\end{cases}\end{split}\\
\label{res2}&\mathcal{L}_{yy}\mean{S_{yy}}+\mathcal{L}_{vv}\mean{S_{vv}}\le \mu^U(\mathcal{L}_{yy}S_{yy}+\mathcal{L}_{vv}S_{vv}),\\
\label{res3}\begin{split}\begin{cases}
&\mathcal{L}_{vx}\mean{T_{vx}}+\mathcal{L}_{xv}\mean{T_{xv}}\ge \mu^L(\mathcal{L}_{vx}T_{vx}+\mathcal{L}_{xv}T_{xv}),\\
&\mathcal{L}_{xx}\mean{T_{xx}}+\mathcal{L}_{vv}\mean{T_{vv}}\le \mu^U(\mathcal{L}_{xx}T_{xx}+\mathcal{L}_{vv}T_{vv}),
\end{cases}\end{split}
\end{align}
where $\mathcal{L}_{lr}=p_lp_rN_t$ is the total pulse pairs sent out by source $lr$.

With those preparations, we can now calculate the secure final key rate. We first calculate the lower bound of $S^+$ with Eqs.~\eqref{splus},~\eqref{slr}, and~\eqref{res1} and the upper bound of $S^-$ with Eqs.~\eqref{sminus},~\eqref{slr}, and~\eqref{res2}.
And then we calculate the upper and lower bounds of $\mathcal{H}$ with Eqs.~\eqref{hu},~\eqref{hl},~\eqref{slr},~\eqref{txx}, and~\eqref{res3}. For each certain $\mathcal{H}$ $(\mathcal{H}\in[\mathcal{H}^L,\mathcal{H}^U])$, we could get one corresponding $\mathcal{R}(\mathcal{H})$ with Eqs.~\eqref{s11},~\eqref{e11}, and~\eqref{RH}. Finally, we get the secure final key rate $R=min[\mathcal{R}(\mathcal{H})]$.
\section{Numerical simulation}\label{simulation}
In this section, we present some numerical simulations to show the efforts of intensity fluctuations. Firstly, we shall estimate what values would be probably observed for the yields and error yields in the normal cases by the linear models~\cite{a8}.We focus on the symmetric case where the two channel transmissions from Alice to UTP and from Bob to UTP are equal. We also assume that the UTP's detectors are identical, i.e., they have the same dark count rates and detection efficiencies, and their detection efficiencies do not depend on the incoming signals. As discussed before, we know that the methods presented in this paper apply to any sources that satisfy the condition given by Eqs.~\eqref{decoycondition1} and~\eqref{decoycondition2}. For simplicity, we suppose that Alice and Bob use weak coherent states (WCS). The density matrix of the WCS with intensity $\mu$ can be written into $\rho=\sum_{k=0}^\infty \frac{e^{-\mu}\mu^{k}}{k!}\oprod{k}{k}$. The actual intensity of the $i$th pulse for source $l$ out of Alice's (or Bob's) laboratory is
\begin{equation}
  \mu_{l}^{i}=\mu_{l}(1+\delta_{l}^{i}), \quad (l=x,y,z),
\end{equation}
and $\mu_v^i\leq \delta_1$ with the boundary conditions $|\delta_{l}^{i}|\leq \delta_2$ for $l=x,y,z$.

\begin{figure}
\centering
\includegraphics[width=8cm]{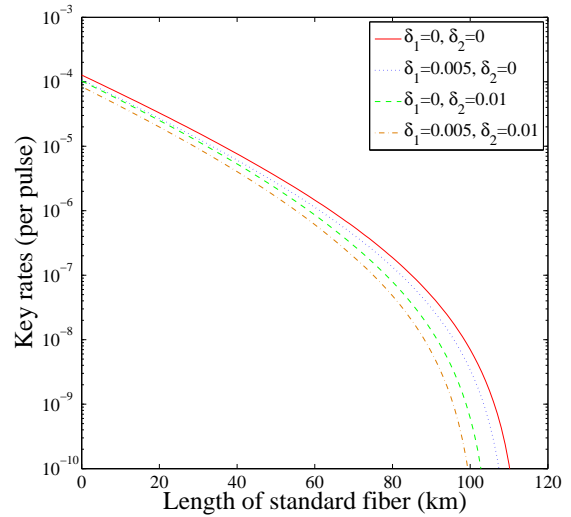}
\caption{(color online) The optimal key rates (per pulse) versus transmission distance (the distance between Alice and Bob) with different intensity fluctuations under the experimental parameters listed in Table \ref{exproperty}. Here we set the total number of pulses at each side $N_t=10^{11}$.}\label{figure1}
\end{figure}

\begin{figure}
\centering
\includegraphics[width=8cm]{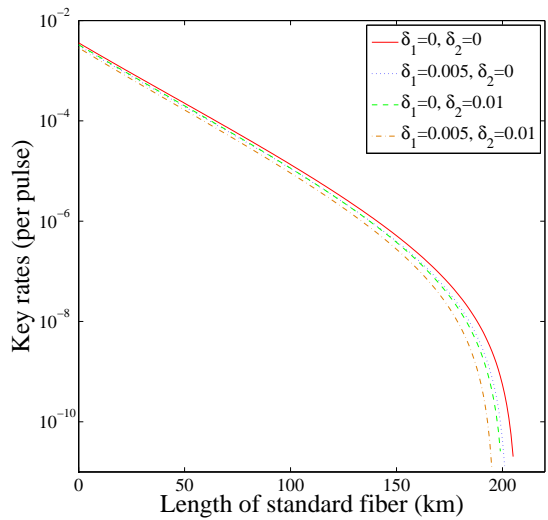}
\caption{(color online) The optimal key rates (per pulse) versus transmission distance (the distance between Alice and Bob) with different intensity fluctuations under the experimental parameters listed in Table \ref{exproperty}. Here we set the total number of pulses at each side $N_t=10^{13}$ and the detection efficiency $\eta_d=40.0\%$.}\label{figure2}
\end{figure}

The values of parameters used in the numerical simulations are listed in Table~\ref{exproperty}. Figures \ref{figure1} and \ref{figure2} 
show the key rates versus transmission distance with different intensity fluctuations. We set the total pulse pairs $N_t=10^{11}$ in Figure \ref{figure1} and $N_t=10^{13}$ and the detection efficiency $\eta_d=40.0\%$ in Figure \ref{figure2}. Figures \ref{figure1} and \ref{figure2} show that the effect of the intensity fluctuation is worse with worse experimental conditions (including smaller data size and worse detectors).

\begin{table}
\begin{ruledtabular}
\begin{tabular}{ccccccc}
$e_0$ & $e_d$ & $p_d$ & $\eta_d$ & $f$ & $\alpha_f$& $\xi$ \\
\hline
0.5 & $1.5\%$ & $6.02\times 10^{-6}$ & $14.5\%$ & $1.16$ & $0.2$& $1.0\times 10^{-7}$ \\ 
\end{tabular}
\end{ruledtabular}
\caption{List of experimental parameters used in numerical simulations. Here $e_0$ is error rate of the vacuum count, $e_d$ is the misalignment-error probability, $p_d$ is the dark count rate of the UTP's detectors, $\eta_d$ is the detection efficiency of the UTP's detectors, $f$ is the error correction inefficiency, $\alpha_f$ is the fiber loss coefficient ($dB/km$), and $\xi$ is the fixed failure probability.}\label{exproperty}
\end{table}

\section{conclusion}\label{conclusion}
In summary, we have shown how to calculate the secure final key rate in the decoy-state MDI-QKD protocol with both source errors and statistical fluctuations with a certain failure probability, provided that the parameters in the diagonal state of the source satisfy Eqs.~\eqref{decoycondition1} and~\eqref{decoycondition2} and bound values of each parameter in the state are known. By our method, all imperfections in a finite pulse pair size decoy-state MDI-QKD protocol with an unstable source have been taken into consideration. Our result can immediately apply to all existing experimental results.

{\bf{Acknowledgement:}} We thank Xiao-Long Hu and Yi-Heng Zhou for discussions. We acknowledge the financial support in part by the 10000-Plan of Shandong province (Taishan Scholars); National High-Tech Program of China Grants No. 2011AA010800 and No. 2011AA010803; National Natural Science Foundation of China Grants No. 11474182, No. 11174177, and No. 60725416; Open Research Fund Program of the State Key Laboratory of Low-Dimensional Quantum Physics Grant No. KF201513; and Key Research and Development Plan Project of ShanDong Province Grant No. 2015GGX101035.

\end{document}